\documentclass[10pt,twocolumn,twoside]{IEEEtran}
\usepackage{amsbsy,amsmath,amsfonts,amssymb,amsbsy,subfigure,verbatim}
\usepackage{bm,cite,graphicx,psfrag,pstricks,theorem,tikz,times,url,verbatim}
\usetikzlibrary{shapes,snakes,calendar,matrix,backgrounds,folding}
\usepackage[english]{babel}
\usepackage[ruled,vlined]{algorithm2e}
\usetikzlibrary{arrows,automata}
\usetikzlibrary{calc} 
\usetikzlibrary{decorations.pathmorphing} 

\newcommand{\bi}{\begin{itemize}}
\newcommand{\ei}{\end{itemize}}
\newcommand{\ben}{\begin{enumerate}}
\newcommand{\een}{\end{enumerate}}

\newcommand{\bc}{\begin{cases}}
\newcommand{\ec}{\end{cases}}
\newcommand{\bd}{\begin{description}}
\newcommand{\ed}{\end{description}}

\newcommand{\be}{\begin{equation}}
\newcommand{\ee}{\end{equation}}
\newcommand{\bea}{\begin{eqnarray}}
\newcommand{\eea}{\end{eqnarray}}

\newtheorem{lemma}{Lemma}

\newtheorem{theorem}{Theorem}

\theoremstyle{plain}
\newtheorem{remark}{Remark}

\graphicspath{%
     {./Figures/}
}

\begin{document}

\title{A new result of the scaling law of weighted $\ell_1$ minimization}

\author{Jun~Zhang,
        Urbashi~Mitra,
        Kuan-Wen~Huang
        and~Nicolo~Michelusi
\thanks{J. Zhang is with College of Information Engineering, Guangdong
University of Technology, Panyu District, Guangzhou 510006, China, and also with the Dept. of Electrical Engineering, University of Southern California, LA, 90089, United States. U. Mitra, K. W. Huang and N. Michelusi are with the Dept. of Electrical Engineering, University of Southern California, LA, 90089, United States. email: jzhang@gdut.edu.cn, ubli@usc.edu.}
}

\maketitle


\begin{abstract}
This paper study recovery conditions of weighted $\ell_1$ minimization for signal reconstruction from compressed sensing measurements.
A sufficient condition for exact recovery by using the general weighted $\ell_1$ minimization is derived, which builds a direct relationship between the weights and the recoverability. Simulation results indicates that this sufficient condition provides a precise prediction of the scaling law for the weighted $\ell_1$ minimization.
\end{abstract}

\section{Introduction}
To recover vector $\boldsymbol{x^*}$ from measurement
\begin{equation}
\bf \boldsymbol{y}=A\boldsymbol{x^*}+\boldsymbol{Z}
\end{equation}
where ${\bf{A}}={\left[\boldsymbol{A}_1, \boldsymbol{A}_2, ..., \boldsymbol{A}_n\right]} \in R^{m \times n}$ is the i.i.d. Gaussian random matrix with rows $\boldsymbol{A}^i$ $\sim \mathcal{N}(0, \sigma_A^2\boldsymbol{I})$ (assume $\boldsymbol{I}$ is the $n \times n$ identity matrix) and ${\boldsymbol{Z}} \sim \mathcal{N}(0, \sigma_Z^2\boldsymbol{I})$ is i.i.d. Gaussian noise, we consider the following weighted $\ell_1$ minimization:
\begin{equation}\label{wl1m}
{\boldsymbol{\hat x}}=\arg\mathop {\min }\limits_{\boldsymbol{x}} \frac{1}{2m}\left\| {\bf {A\boldsymbol{x}} - \boldsymbol{y}} \right\|_2^2 + h{\left\| {\bf {W\boldsymbol{x}}} \right\|_1},\;h > 0
\end{equation}
where ${\bf{W}} \in R^{n \times n}$ whose off-diagonal elements are zero and the diagonal elements
\begin{equation}\label{weights}
{w_i} \in (0,\; + \infty )
\end{equation}
Note that due to the present of noise, it is generally impossible to seek exact recovery of the sparse signal $\boldsymbol{x^*}$. Accordingly, this paper focuses on the goal that the optimum solution $\boldsymbol{\hat x}$ and the true signal $\boldsymbol{x^*}$ have their nonzero entries at the same locations and with same signs, i.e., \emph{sparsity pattern} recovery or \emph{support} recovery.

\section{Main Results}

At first, we denote the subdifferential of $\left\| {\bf{W\boldsymbol{x}}} \right\|_1$ as
\begin{equation}\label{subd}
\begin{split}
\partial {\left\| {\bf W\boldsymbol{x}} \right\|_1} &= \{ \bf{W\boldsymbol{u}}|{\boldsymbol{u}^T}{\bf W}\boldsymbol{x} = {\left\| {\bf W\boldsymbol{x}} \right\|_1},\;{\left\| \boldsymbol{u} \right\|_\infty } \le 1\}\\
&= \{ {\bf W\boldsymbol{u}}|{u_i} = \mbox{sign}({x_i}),\;\text{if}\;{x_i} \ne 0\;\\
&\;\;\;\;\;\;\;\;\;\;\;\;\;\;\;\;\;\;\;\;\text{and}\; {{u_i}} \in [-1, 1],\;\text{otherwise}\}
\end{split}
\end{equation}
where
\begin{equation}
\mbox{sign}({x_i}) \buildrel \Delta \over = \left\{ \begin{array}{l}
 + 1,\quad \text{if}\;{x_i} > 0\\
 - 1,\quad \text{if}\;{x_i} < 0\\
0,\quad \;\;\;\text{if}\;{x_i} = 0
\end{array} \right.
\end{equation}
From convex analysis, we introduce the following Lemma.
\begin{lemma}\label{scl}
(a) A vector $\boldsymbol{\hat x} \in R^n$ is a global minimum of the model (\ref{wl1m}) if and only if $\exists \,\bf{W\boldsymbol{\hat u}} \in \partial {\left\| {\bf W\boldsymbol{x}} \right\|_1},\text{such that}$
\begin{equation}\label{SNC}
\frac{1}{m}{{\bf A^T}}(\bf{A{\boldsymbol{\hat x}}} - \boldsymbol{y}) + h{\bf W\boldsymbol{\hat u}} = \boldsymbol{0}
\end{equation}
(b) If $|\hat u_i| < 1$ for all $i \notin \boldsymbol{\hat S}$ and $\bf A_{\boldsymbol{\hat S}}$ is full rank, then $\boldsymbol{\hat x}$ is the unique minimum and $\hat x_i=0$ for all $i \notin \boldsymbol{\hat S}$ where $\boldsymbol{\hat S}$ denotes the support set of vector $\boldsymbol{\hat x}$.
\end{lemma}
The proof of this Lemma is given in Appendix A.

\begin{remark}
Because measurement matrix $\bf A$ is Gaussian random matrix which has full rank with a probability of one, we suppose the condition $\bf A_{\boldsymbol{\hat S}}$ (or $\bf A_{\boldsymbol{S}}$) full rank is always satisfied throughout this paper. Note that for a matrix (or vector) $\bf M$, we denote $\bf M_{\boldsymbol{\Lambda}}$ the reduced dimensional matrix (or vector) built upon the columns (or entries) of $\bf M$ whose indices are included in set $\boldsymbol{\Lambda}$.
\end{remark}

Therefore, $\boldsymbol{\hat x}$ is the unique minimum of model (\ref{wl1m}) if
\begin{equation}\label{SNC2}
\begin{split}
&\frac{1}{m}{{{\bf{A}}}^T_{\boldsymbol{\hat S}}}({\boldsymbol{y}} - {\bf{A_{\boldsymbol{\hat S}}}{{\boldsymbol{\hat x}}_{\boldsymbol{\hat S}}}})=h\cdot{\bf{W_{\boldsymbol{\hat S}}}}\cdot{\boldsymbol{\hat u_{\hat S}}}\\
&\left| {\frac{\boldsymbol{A}_i^T}{m}({\boldsymbol{y}} - {\bf{A_{\boldsymbol{\hat S}}}{{\boldsymbol{\hat x}}_{\boldsymbol{\hat S}}}})} \right| < hw_i\quad for\;i \notin \boldsymbol{\hat S}
\end{split}
\end{equation}
On the other hand, assume $\boldsymbol{S}$ is the support set of the true signal $\boldsymbol {x^*}$ with cardinality $|{\boldsymbol{S}}|=k\ll n$. Denote $\boldsymbol{S^c}=[1,2,...,n]\setminus \boldsymbol{S}$ where $\setminus$ represents set difference. We can establish a sufficient condition under which the model (\ref{wl1m}) recovers its support exactly, i.e., $\text{sign}(\boldsymbol{\hat x})=\text{sign}(\boldsymbol{x^*})$.
\begin{lemma}\label{SC}
The support of signal $\boldsymbol{x^*}$ can be recovered exactly from the solution of model (\ref{wl1m}), i.e., $\mbox{sign}(\boldsymbol{\hat x})=\mbox{sign}(\boldsymbol{x^*})$, provided the following events are satisfied
\begin{equation}\label{SNC3}
\begin{split}
&1)\;\left| {\frac{{{\boldsymbol{A}}_i^T}}{m}\left\{ {(\boldsymbol{I} - {{\bf{A}}_{\boldsymbol{S}}}{\bf{A}}_{\boldsymbol{S}}^ + )\boldsymbol{Z} + mh{{\bf{A}}_{\boldsymbol{S}}}{{({\bf{A}}_{\boldsymbol{S}}^T{{\bf{A}}_{\boldsymbol{S}}})}^{ - 1}}{\bf W_{\boldsymbol{S}}}{\boldsymbol{ u}_{\boldsymbol{S}}}} \right\}} \right| \\
&\;\;\;\;\;\;\;\;\;\;\;\;\;\;\;\;\;\;\;\;\;\;\;\;\;\;\;\;\;\;\;\;\;\;\;\;\;\;\;\;\;\;\;\;\;\;\;\;\;\;\;\;\;\;\;\;\;\;< h{w_i},\quad \forall \;i \in {{\boldsymbol{S}}^{{c}}}\\
&2)\;\mbox{sign}({\boldsymbol{x^*_{S}}}+{{\bf A}^+_{\boldsymbol{S}}\boldsymbol{Z}}-mh{({\bf A}^T_{\boldsymbol{S}}{\bf A}_{\boldsymbol{S}})^{-1}{\bf W}_{\boldsymbol{S}}\boldsymbol{u}_{\boldsymbol{S}}})=\mbox{sign}(\boldsymbol{x^*_{S}})
\end{split}
\end{equation}
where ${\bf{A}}_{\boldsymbol{S}}^ +  = {({\bf{A}}_{\boldsymbol{S}}^T{{\bf{A}}_{\boldsymbol{S}}})^{ - 1}}{\bf{A}}_{\boldsymbol{S}}^T$ is the pseudoinverse of ${\bf{A}}_{\boldsymbol{S}}$ and ${\boldsymbol{ u}_{\boldsymbol{S}}}=\mbox{sign}(\boldsymbol{x^*_{S}})$.
\end{lemma}
The proof of this Lemma is given in Appendix A.

\begin{remark}
The first condition in (\ref{SNC3}) is a recovery guarantee for the zero entries in the signal $\boldsymbol{x^*}$ from which we can find that the recoverability of the zero entries in the signal $\boldsymbol{x^*}$ obtained by solving the problem (\ref{wl1m}) depends on the measurement matrix $\bf A$, the weights $\bf W$, the parameter $h$, the noise $\boldsymbol{Z}$ and the sign pattern of signal $\boldsymbol{x^*}$, but not on the magnitudes of its nonzero entries. Whereas, from the second condition in (\ref{SNC3}), the recoverability of the nonzero entries in the signal $\boldsymbol{x^*}$ is related with its magnitudes besides the factors mentioned above.
\end{remark}

Next, precise conditions on the system parameters $(m, n, k)$ can be obtained which are sufficient to guarantee the support recovery. We state the conclusion in the following Theorem.
\begin{theorem}\label{recoverytheorem}
For an $k$-sparse signal $\boldsymbol{x^*}$ with $k \ll n$, problem (\ref{wl1m}) is solved to recover its support $\boldsymbol{S}$ from linear measurement $\bf \boldsymbol{y}=A\boldsymbol{x^*}+\boldsymbol{Z}$. Define the gap
\begin{equation}
g(h)\buildrel \Delta \over ={c_3}h{\left\| {{{\bf{W}}_{\boldsymbol{S}}}{{\boldsymbol{u}}_{\boldsymbol{S}}}} \right\|_\infty } + 6\sqrt {\frac{{\sigma _Z^2\log (k)}}{{m\sigma _A^2}}}
\end{equation}
If $\forall i \in \boldsymbol{S}$ $|x_i^*|>g(h)$ holds, and if for some fixed $\epsilon^{'}>0$, triple (m, n, k) and regularization parameter $h$ obeys
\begin{equation}\label{threshold}
m > 2\eta k\log (n - k)(1 + \epsilon^{'})\left(1 + \frac{{\sigma _Z^2\sigma _A^2}}{{{h^2}k}}\right)
\end{equation}
where $\eta=\mathop {\max}\limits_{i \in \boldsymbol{S}^c}\left\{\frac{\xi}{w_i^2}\right\}$ with $\xi=\frac{\sum\limits_{i = 1}^k {W_{S,i}^2}}{k}$ and $W_{S,i}$ represents the $i$-th diagonal element in the matrix $\bf W_{\boldsymbol{S}}$, then the solution $\boldsymbol{\hat x}$ of problem (\ref{wl1m}), with probability greater
than 1-$c_1${exp}($-c_2${min}$\{k$, $log(n-k)\})$ for some positive constants $c_1$ and $c_2$, recovers the support of signal $\boldsymbol{x^*}$ exactly, i.e., $\text{sign}(\boldsymbol{\hat x})=\text{sign}(\boldsymbol{x^*})$.
\end{theorem}
The proof of this Theorem is given in Appendix B.

\begin{remark}
Theorem \ref{recoverytheorem} indicates that if $m>2\eta klog(n-k)$ holds and the nonzero entries of $\boldsymbol{x^*}$ are large enough, model (\ref{wl1m}) can, with high probability, recover the support of signal $\boldsymbol{x^*}$ exactly where the important parameters $\eta$ are directly related to the model weights. In real applications, we can significantly reduce the sample requirement for support recovery through optimizing the weights in model (\ref{wl1m}) so as to achieve the $\eta$ as small as possible.
\end{remark}

\begin{remark}\label{hset}
A result similar to the one in \cite{wainwright2009sharp} can be shown, if we set
\begin{equation}\label{hequ}
h=\sqrt {\frac{{{2\phi _n}\eta\sigma _Z^2\sigma _A^2\log (n-k)}}{m}}
\end{equation}
for some $\phi _n\geq2$, then it suffices to have $m>2\eta k\log (n - k)\left(({1+\epsilon^{'}})^{-1}-\frac{1}{\phi _n}\right)^{-1}$ for some $\epsilon^{'}>0$. Moreover, if we choose an $h$ with $\phi _n\rightarrow +\infty$, then Theorem \ref{recoverytheorem} guarantees the support recovery of $\boldsymbol{x^*}$ with about $m=2\eta k\log (n - k)$ samples.
\end{remark}

\begin{remark}
A special case of weighted $\ell_1$ minimization model is the Modified-CS \cite{Va_MCS} which weights the partial known support as zero. According to condition (\ref{threshold}), we can find that if the prior support information is accuracy, this weight strategy ensures that $\eta<1$ holds. Comparing with the classical result $m>2klog(n-k)$ \cite{wainwright2009sharp} required by the BPDN where $\eta=1$, Modified-CS achieves a reduced sample requirement by exploiting prior support information.
\end{remark}

\begin{figure}
  \centering
  \subfigure[]{
    \label{fig1:subfig:a} 
    \includegraphics[width=3in, height=2.4in]{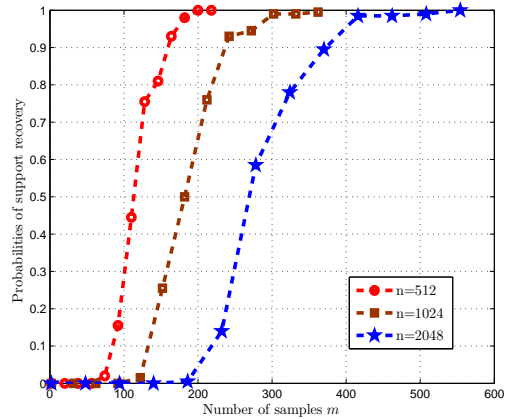}
  }
  \subfigure[]{
    \label{fig1:subfig:b} 
    \includegraphics[width=3in, height=2.4in]{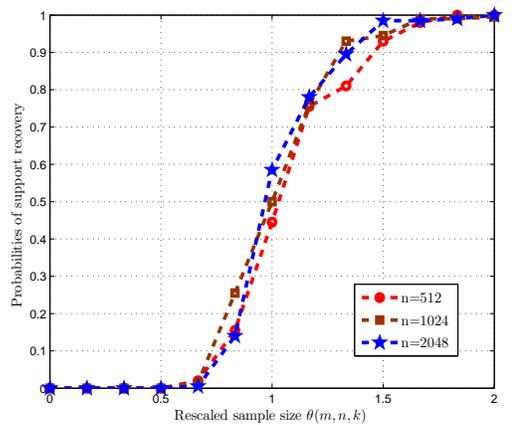}
  }
  \caption{Simulation results of model (\ref{wl1m}) with $\eta=1$. (a) The probabilities of support recovery versus the sample size $m$ for three different problem size $n$, in all cases with sparsity $k=\left\lceil {0.4{n^{0.5}}} \right\rceil$. (b) The probabilities of support recovery versus the rescaled sample size $\theta(m,n,k)=m/[2klog(n-k)]$.}
  \label{fig1:subfig} 
\end{figure}

\section{Simulation Results}
In this section, some simulations have been conducted to validate the scaling law built in Theorem \ref{recoverytheorem}. In our experiments, the nonzero element of $k$-sparse signals is $\pm1$ uniformly at random. Measurement matrix $\textbf{A} \in \mathbb{R}^{m \times n}$ is drawn randomly from the standard Guassian distribution, i.e., ${A_{i,j}} \sim i.i.d\; \mathcal{N}(0,1)$, and noise $\boldsymbol{Z} \sim \mathcal{N}(0,{\sigma_Z ^2}\bf{I})$ with $\sigma_Z=0.5$. Based on Remark \ref{hset}, the choice of $h$ follows equation (\ref{hequ}) with $\phi _n=9$ in our experiments. At first, the standard BP model is employed to recover the support of the $k$-sparse signals $\bf x^*$. According to Theorem \ref{recoverytheorem}, the standard BP model, as a special case of the weighted $\ell_1$ minimization model, has $\eta=1$. In Fig. \ref{fig1:subfig:a}, we plot the probabilities of support recovery versus the sample size $m$ for three different problem sizes $n \in \{ 512,1024,2048\}$, and $k=\left\lceil {0.4{n^{0.5}}} \right\rceil$ in each case. We repeat each experiment 200 times at each point. Obviously, the probabilities of support recovery vary from zero to one along with the samples increase and the larger problem requires more samples. However, according to the scaling predicted by Theorem \ref{recoverytheorem}, i.e.,
\begin{equation}
m > 2\eta k\log (n - k)(1 + \epsilon)(1 + \frac{{\sigma _Z^2\sigma _A^2}}{{{h^2}k}})\buildrel \Delta \over = 2\eta \zeta k\log (n - k)
\end{equation}
where $\zeta$ is a constant. Thus, Fig. \ref{fig1:subfig:b} plots the same experimental results but the probabilities of support recovery are now plotted versus an ``appropriately rescaled" version of the sample size, i.e., $\theta(m,n,k)=m/[2klog(n-k)]$. In Fig. \ref{fig1:subfig:b}, all of the curves now line up with one another, even though the problem sizes and sparsity levels vary dramatically. And all of the cases obtain the probabilities of support recovery are equal to one at $\theta(m,n,k)=\zeta \approx 2$. Obviously, the experimental result matches the theoretical prediction in Theorem \ref{recoverytheorem} very well. Note that similar simulation was carried out in \cite{wainwright2009sharp} to confirm the scaling law of standard BP model.

\begin{figure}
  \centering
  \subfigure[]{
    \label{fig2:subfig:a} 
    \includegraphics[width=3in, height=2.4in]{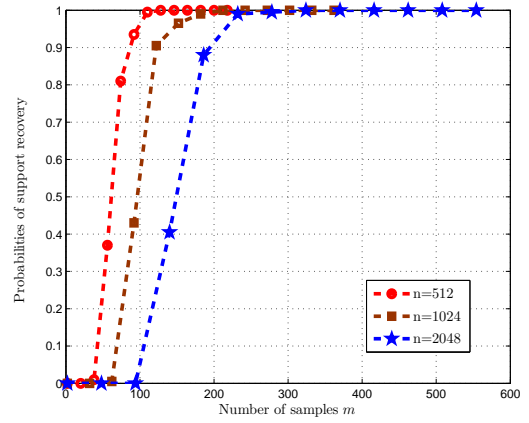}
  }
  \subfigure[]{
    \label{fig2:subfig:b} 
    \includegraphics[width=3in, height=2.4in]{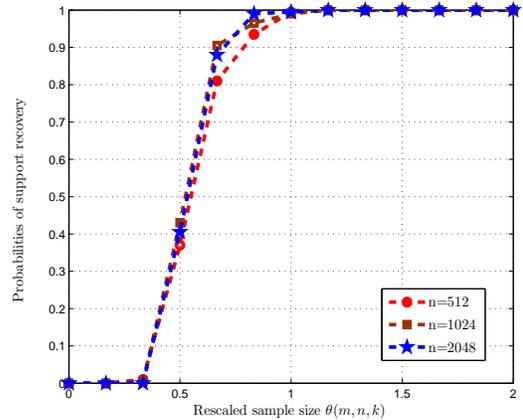}
  }
  \caption{Simulation results of model (\ref{wl1m}) with $\eta=0.5$. (a) The probabilities of support recovery versus the sample size $m$ for three different problem size $n$, in all cases with sparsity $k=\left\lceil {0.4{n^{0.5}}} \right\rceil$. (b) The probabilities of support recovery versus the rescaled sample size $\theta(m,n,k)=m/[2klog(n-k)]$.}
  \label{fig2:subfig} 
\end{figure}

\begin{figure}
  \centering
  \subfigure[]{
    \label{fig3:subfig:a} 
    \includegraphics[width=3in, height=2.4in]{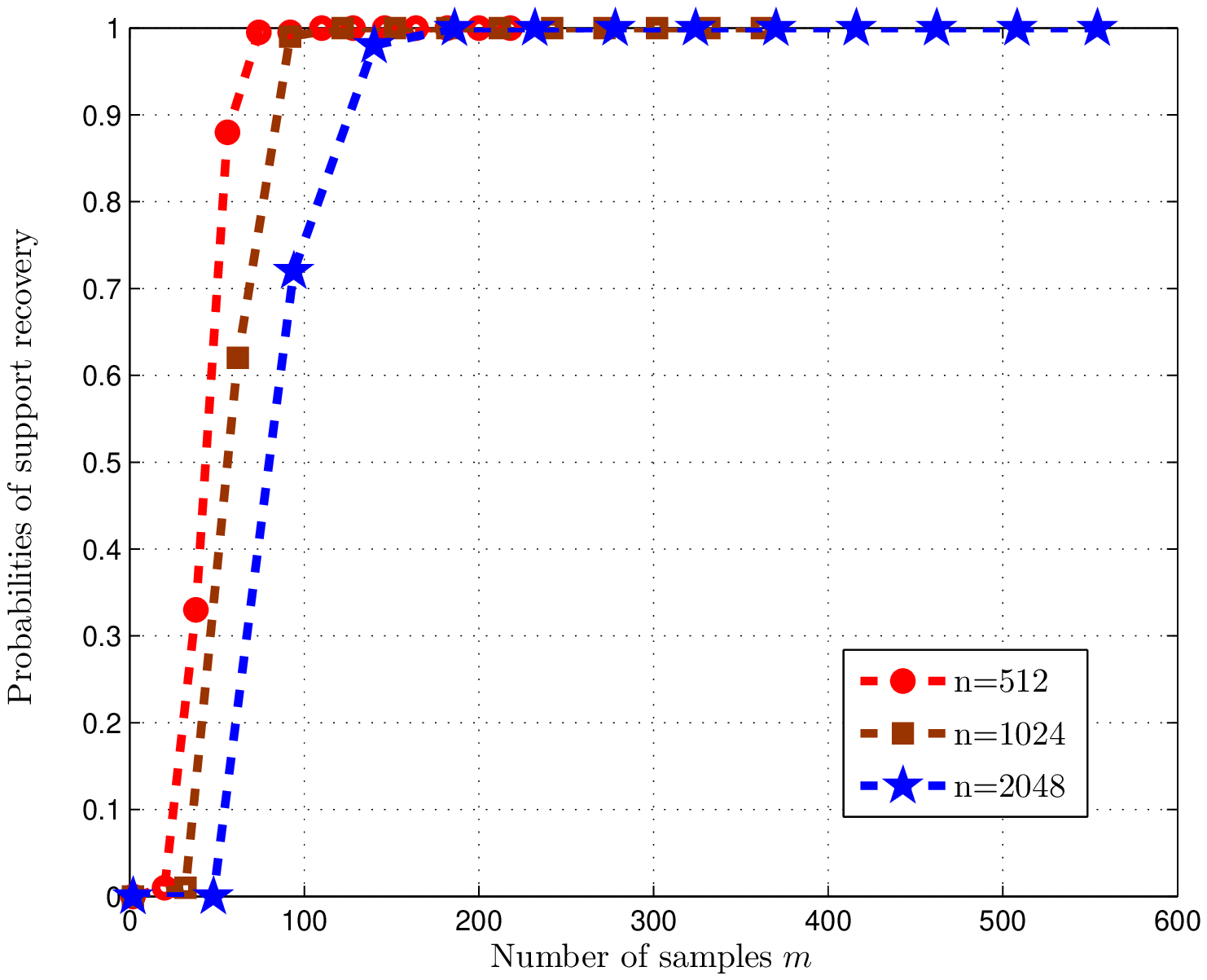}
  }
  \subfigure[]{
    \label{fig3:subfig:b} 
    \includegraphics[width=3in, height=2.4in]{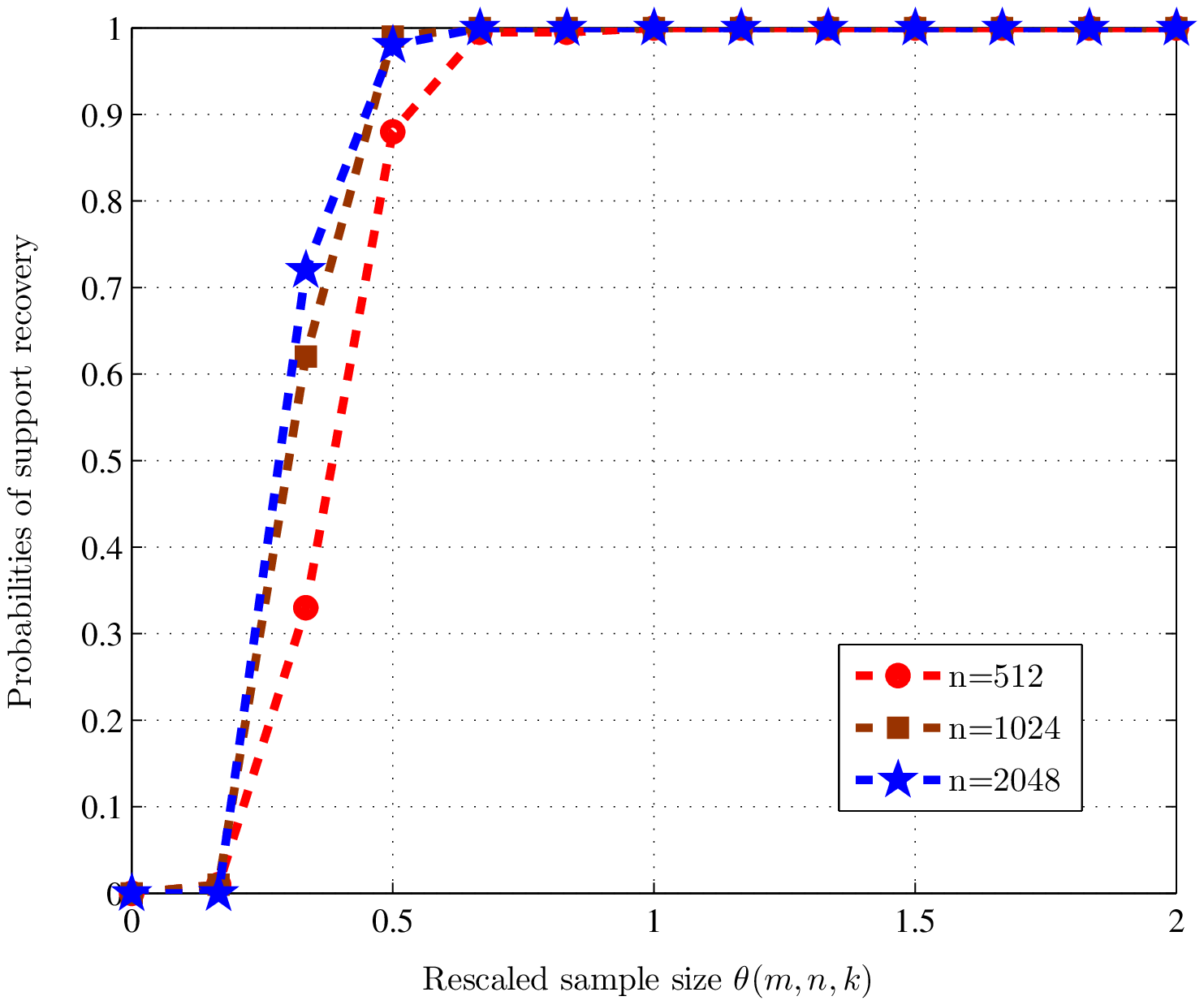}
  }
  \caption{Simulation results of model (\ref{wl1m}) with $\eta=0.25$. (a) The probabilities of support recovery versus the sample size $m$ for three different problem size $n$, in all cases with sparsity $k=\left\lceil {0.4{n^{0.5}}} \right\rceil$. (b) The probabilities of support recovery versus the rescaled sample size $\theta(m,n,k)=m/[2klog(n-k)]$.}
  \label{fig3:subfig} 
\end{figure}

Further, the same experiments are performed but we used the weighted $\ell_1$ minimization model where the weights aren't equal to one to recover the support of the $k$-sparse signal $\boldsymbol{x^*}$. Two classes of weights are tested where one weights nonzero element of $k$-sparse signals with $w_i=\sqrt 2/2$ and another is $w_i=1/2$. The experimental results are plotted in Fig. \ref{fig2:subfig} and Fig. \ref{fig3:subfig}, respectively. According to Theorem \ref{recoverytheorem}, the weighted $\ell_1$ minimization model have $\eta=0.5$ and $\eta=0.25$ with respective to the two classes of weights respectively. As shown in Fig. \ref{fig2:subfig} and Fig. \ref{fig3:subfig}, the curves obtain the probabilities of support recovery are equal to one at $\theta(m,n,k)=\frac{1}{2}\zeta \approx 1$ and $\theta(m,n,k)=\frac{1}{4}\zeta \approx 0.5$, respectively. Obviously, all of the simulation results match the theoretical predictions in Theorem \ref{recoverytheorem} very well, which indicates that Theorem \ref{recoverytheorem} provides a precise prediction of the scaling law for the weighted $\ell_1$ minimization.

\section*{APPENDIX A}
\subsection{Proof of Lemma 1}

\emph{Proof:} It is well known that the problem (\ref{wl1m}) can be transferred into an equivalent constrained problem that involves a continuous objective function over a compact set \cite{Luenberger}. Therefore, its minimum is always achieved. Based on the first order optimality condition \cite{HiriartC}, $\boldsymbol{\hat x}$ is a global minimum for the model (\ref{wl1m}) if and only if $\exists \,{\bf W}\boldsymbol{\hat u} \in \partial {\left\| {\bf W}\boldsymbol{x} \right\|_1},\text{such that}$ $\frac{1}{m}{{\bf A}^T}({\bf A}\boldsymbol{{\hat x}} - \boldsymbol{y}) + h{\bf W}\boldsymbol{\hat u} = \bf 0$. Thereby Lemma 1(a) is established.

According to the standard duality theory \cite{Luenberger}, given the subgradient ${\bf W}\boldsymbol{\hat u} \in R^n$, any optimum $\boldsymbol{\hat x} \in R^n$ of model (\ref{wl1m}) must satisfy the complementary slackness condition ${\boldsymbol{\hat u}^T}{\bf W}\boldsymbol{\hat x} = {\left\| {\bf W\boldsymbol{\hat x}} \right\|_1}$. For all $i$ such that $|\hat u_i|<1$, this condition holds if and only if $\hat x_i=0$. Further, if $|\hat u_i| < 1$ for all $i \notin \boldsymbol{\hat S}$ and $\bf A_{\boldsymbol{\hat S}}$ is full rank, then $\boldsymbol{\hat x}$ can be determined uniquely from (\ref{SNC}). Therefore, Lemma 1(b) holds.

\subsection{Proof of Lemma 2}

\emph{Proof:} Define an $n$-dimensional vector $\boldsymbol{x^\dag}$ as
\begin{equation}\label{eq14}
\left\{ \begin{array}{l}
{\boldsymbol{x}}_{\boldsymbol{S}}^\dag  = {\boldsymbol{x}}_{\boldsymbol{S}}^* + {\bf{A}}_{\boldsymbol{S}}^ + {\boldsymbol{Z}} - mh{({\bf{A}}_{\boldsymbol{S}}^T{{\bf{A}}_{\boldsymbol{S}}})^{ - 1}}{{\bf{W}}_{\boldsymbol{S}}}{{\boldsymbol{u}}_{\boldsymbol{S}}}\\
{\boldsymbol{x}}_{{{\boldsymbol{S}}^c}}^\dag  = {\boldsymbol{0}}
\end{array} \right.
\end{equation}
where ${{\boldsymbol{u}}_{\boldsymbol{S}}}=\mbox{sign}({\boldsymbol{x}}_{\boldsymbol{S}}^*)$. If the conditions in (\ref{SNC3}) are satisfied, we will prove that vector $\boldsymbol{x^\dag}$ is the unique minimum of model (\ref{wl1m}).

According to the second condition in (\ref{SNC3}), we have
\begin{equation}\label{eq15}
\mbox{sign}({\boldsymbol{x}}_{\boldsymbol{S}}^\dag)=\mbox{sign}({\boldsymbol{x}}_{\boldsymbol{S}}^*)
\end{equation}
At the same time, utilizing the equality ${{\boldsymbol{u}}_{\boldsymbol{S}}}=\mbox{sign}({\boldsymbol{x}}_{\boldsymbol{S}}^*)$, it follows
\begin{equation}\label{eq12}
{\boldsymbol{x}}_{\boldsymbol{S}}^\dag  = {\boldsymbol{x}}_{\boldsymbol{S}}^* + {\bf{A}}_{\boldsymbol{S}}^ + {\boldsymbol{Z}} - mh{({\bf{A}}_{\boldsymbol{S}}^T{{\bf{A}}_{\boldsymbol{S}}})^{ - 1}}{{\bf{W}}_{\boldsymbol{S}}}\times\mbox{sign}({\boldsymbol{x}}_{\boldsymbol{S}}^\dag)
\end{equation}
Obviously, ${\boldsymbol{x}}_{\boldsymbol{S}}^\dag$ satisfies the first condition in (\ref{SNC2}) with $\boldsymbol{S}$ replaced by $\boldsymbol{\hat S}$ and ${\boldsymbol{x}}_{\boldsymbol{S}}^\dag$ replaced by ${\boldsymbol{\hat x}}_{\boldsymbol{\hat S}}$.

Further, substituting (\ref{eq12}) into the second condition in (\ref{SNC2}), we have that $\forall i \notin \boldsymbol{S}$
\begin{equation}\label{eq13}
\begin{split}
&\left| {\frac{\boldsymbol{A}_i^T}{m}({\boldsymbol{y}} - {\bf{A_{\boldsymbol{S}}}{{\boldsymbol{x}}_{\boldsymbol{S}}^{\dag}}})} \right|\\
&=\left| {\frac{{{\boldsymbol{A}}_i^T}}{m}\left\{ {(\boldsymbol{I} - {{\bf{A}}_{\boldsymbol{S}}}{\bf{A}}_{\boldsymbol{S}}^ + )\boldsymbol{Z} + mh{{\bf{A}}_{\boldsymbol{S}}}{{({\bf{A}}_{\boldsymbol{S}}^T{{\bf{A}}_{\boldsymbol{S}}})}^{ - 1}}{\bf W_{\boldsymbol{S}}}\times\mbox{sign}({\boldsymbol{x}}_{\boldsymbol{S}}^\dag)} \right\}} \right|\\
&<hw_i
\end{split}
\end{equation}
where the inequality in (\ref{eq13}) utilizes the fact that $\mbox{sign}({\boldsymbol{x}}_{\boldsymbol{S}}^\dag)={{\boldsymbol{u}}_{\boldsymbol{S}}}$ and follows from the first condition in (\ref{SNC3}). Hence, According to the sufficient conditions in (\ref{SNC2}), $\boldsymbol{x^\dag}$ is the unique minimum of model (\ref{wl1m}), i.e., $\boldsymbol{\hat x}=\boldsymbol{x^\dag}$. Based on (\ref{eq14}) and (\ref{eq15}), $\mbox{sign}(\boldsymbol{\hat x})=\mbox{sign}(\boldsymbol{x^*})$ holds.

\section*{APPENDIX B}
\emph{Proof of Theorem 1}
In this section, the proof of Theorem 1 uses the
techniques from \cite{wainwright2009sharp}, with appropriate modification to account
for the weighted $\ell_1$ norm that replaces the $\ell_1$ norm.

\emph{Proof:}
Based on Lemma \ref{SC}, we conclude that model (\ref{wl1m}) can recover the support of $\boldsymbol{x^*}$ exactly, provided the events in (\ref{SNC3}) are satisfied. Therefore, we firstly will derive a precise condition under which event 1) in (\ref{SNC3}) is satisfied with high probability. Further, by bounding the quantity $({\bf A_{\boldsymbol{S}} ^ + \boldsymbol{Z}} - mh({\bf A_{\boldsymbol{S}}^T{\bf A_{\boldsymbol{S}} }})^{-1}{\bf {W_{\boldsymbol{S}} }{\boldsymbol{u_{S}} }})$, another condition can be obtained to guarantee $\mbox{sign}({\boldsymbol{\hat x_S}})=\mbox{sign}(\boldsymbol{x_s^*})$ holds with high probability. Then, according to Lemma \ref{SC}, the support of signal $\boldsymbol{x^*}$ is, with high probability, recovered exactly from the solution of model (\ref{wl1m}).

For the event 1) in (\ref{SNC3}), conditioned on $\bf A_{\boldsymbol{S}}$ and noise $\boldsymbol{Z}$, we have that
\begin{equation}
{\Gamma _i} \buildrel \Delta \over = \frac{{{\boldsymbol{A_i}^T}}}{{mh}}[({\boldsymbol{I}} - {{\bf{A}}_{\boldsymbol{{S} }}}{\bf{A}}_{\boldsymbol{{S} }}^ + ){\boldsymbol{Z}} + {{mh}}{{\bf{A}}_{\boldsymbol{{S} }}}{({\bf{A}}_{\boldsymbol{{S} }}^{{T}}{{\bf{A}}_{\boldsymbol{{S} }}})^{ - {{1}}}}{{\bf{W}}_{\boldsymbol{{S} }}}{{\boldsymbol{u}}_{\boldsymbol{{S} }}}]
\end{equation}
is zero-mean Gaussian with variance at most
\begin{equation}
\begin{split}
&{\mathop{\rm var}}({\Gamma _i|\bf A_{\boldsymbol{S}},\;\boldsymbol{Z}})\\
&\le \sigma_A^2\left\| {{{\bf{A}}_{\boldsymbol{{S} }}}{{({\bf{A}}_{\boldsymbol{{S} }}^{{T}}{{\bf{A}}_{\boldsymbol{{S} }}})}^{ - {{1}}}}{{\bf{W}}_{\boldsymbol{{S} }}}{{\boldsymbol{u}}_{\boldsymbol{{S} }}} + ({\boldsymbol{I}} - {{\bf{A}}_{\boldsymbol{{S} }}}{\bf{A}}_{\boldsymbol{{S} }}^ + )\frac{{\boldsymbol{Z}}}{{mh}}} \right\|_2^2
\end{split}
\end{equation}
Further, because
\begin{equation}
\left< {{\bf{A}}_{\boldsymbol{{S} }}}{({\bf{A}}_{\boldsymbol{{S} }}^{{T}}{{\bf{A}}_{\boldsymbol{{S} }}})^{ - {{1}}}}{{\bf{W}}_{\boldsymbol{{S} }}}{{\boldsymbol{u}}_{\boldsymbol{{S} }}},\,({\boldsymbol{I}} - {{\bf{A}}_{\boldsymbol{{S} }}}{\bf{A}}_{\boldsymbol{{S} }}^ + )\frac{{\boldsymbol{Z}}}{{mh}} \right>  = 0
\end{equation}
by applying the Pythagorean Theorem, it follows that
\begin{equation}\label{eq2}
\begin{split}
&{\mathop{\rm var}} ({\Gamma _i|\bf A_{\boldsymbol{S}},\;\boldsymbol{Z}}) \\
&\le \sigma_A^2(\left\| {{{\bf{A}}_{\boldsymbol{{S} }}}{{({\bf{A}}_{\boldsymbol{{S} }}^{{T}}{{\bf{A}}_{\boldsymbol{{S}}}})}^{ - {{1}}}}{{\bf{W}}_{\boldsymbol{{S} }}}{{\boldsymbol{u}}_{\boldsymbol{{S} }}}} \right\|_2^2 + \left\| {({\boldsymbol{I}} - {{\bf{A}}_{\boldsymbol{{S} }}}{\bf{A}}_{\boldsymbol{{S} }}^ + )\frac{{\boldsymbol{Z}}}{{mh}}} \right\|_2^2)
\end{split}
\end{equation}
For the first term in equation (\ref{eq2}), we have
\begin{equation}\label{eq3}
\begin{split}
&\left\| {{{\bf{A}}_{\boldsymbol{{S} }}}{{({\bf{A}}_{\boldsymbol{{S} }}^{{T}}{{\bf{A}}_{\boldsymbol{{S} }}})}^{ - {{1}}}}{{\bf{W}}_{\boldsymbol{{S} }}}{{\boldsymbol{u}}_{\boldsymbol{{S} }}}}\right\|_2^2 \quad\quad\quad\quad\quad\quad\quad\quad\quad\quad\quad\quad\\
&\quad\quad\quad\quad\quad= \frac{1}{m}{\boldsymbol{{u}}_{\boldsymbol{S}}^T{\bf W_{\boldsymbol{S}} }}{\left(\frac{{{\bf{A}}_{\boldsymbol{{S} }}^{{T}}{{\bf{A}}_{\boldsymbol{{S} }}}}}{m}\right)^{ - {{1}}}}{{\bf{W}}_{\boldsymbol{{S} }}}{{\boldsymbol{u}}_{\boldsymbol{{S} }}}\\
&\quad\quad\quad\quad\quad\le \frac{1}{m}{\left\| {\boldsymbol{u}_{\boldsymbol{S}}^T{\bf W_{\boldsymbol{S}} }} \right\|_2}{\left\| {{{\left(\frac{{{\bf{A}}_{\boldsymbol{{S} }}^{{T}}{{\bf{A}}_{\boldsymbol{{S} }}}}}{m}\right)}^{ - {{1}}}}{{\bf{W}}_{\boldsymbol{{S} }}}{{\boldsymbol{u}}_{\boldsymbol{{S} }}}} \right\|_2}\\
& \quad\quad\quad\quad\quad\le \frac{1}{m}\left\| {\boldsymbol{u}_{\boldsymbol{S}}^T{\bf W_{\boldsymbol{S}} }} \right\|_2^2{\left| {\left\| {{{\left(\frac{{{\bf{A}}_{\boldsymbol{{S} }}^{{T}}{{\bf{A}}_{\boldsymbol{{S} }}}}}{m}\right)}^{ - {{1}}}}} \right\|} \right|_2}
\end{split}
\end{equation}
where the first inequality follows from the Cauchy-Schwartz inequality, ${\left| {\left\| \cdot \right\|} \right|_2}$ represents the spectral norm and the second inequality follows from the definition of matrix norm.

At the same time, we have
\begin{equation}
\begin{split}
&{\left| {\left\| {{{\left(\frac{{{\bf{A}}_{\boldsymbol{{S} }}^{{T}}{{\bf{A}}_{\boldsymbol{{S} }}}}}{m}\right)}^{ - {{1}}}}} \right\|} \right|_2} \\
&\le {\left| {\left\| {{{(\sigma_A^2\boldsymbol{I} )}^{-1}}} \right\|} \right|_2} + {\left| {\left\| {{{\left(\frac{{{\bf{A}}_{\boldsymbol{{S} }}^{{T}}{{\bf{A}}_{\boldsymbol{{S} }}}}}{m}\right)}^{ - {{1}}}} - {{(\sigma_A^2\boldsymbol{I} )}^{ - 1}}} \right\|} \right|_2}
\end{split}
\end{equation}
Applying Lemma 9 in \cite{wainwright2009sharp}, it follows that event
\begin{equation}\label{eq4}
{\left| {\left\| {{{\left(\frac{{{\bf{A}}_{\boldsymbol{{S} }}^{{T}}{{\bf{A}}_{\boldsymbol{{S} }}}}}{m}\right)}^{ - {{1}}}}} \right\|} \right|_2} \le
\frac{1}{\sigma_A^2}+\frac{8}{\sigma_A^2}\sqrt {\frac{{k}}{m}}
\end{equation}
is satisfied with probability greater than $1-2\text{exp}(-k/2)$.

Recall that the definition of vector $\boldsymbol{u_S}$. We have
\begin{equation}\label{eq5}
\left\| {\boldsymbol{u}_{\boldsymbol{S}}^T{\bf W_{\boldsymbol{S}} }} \right\|_2^2=\sum\limits_{i = 1}^k {W_{S,i}^2}=k\xi
\end{equation}
where $\xi=\frac{\sum\limits_{i = 1}^k {W_{S,i}^2}}{k}$ and $W_{S,i}$ represents the $i$-th diagonal element in the matrix $\bf W_{\boldsymbol{S}}$. Consequently, combining equations (\ref{eq3}), (\ref{eq4}) and (\ref{eq5}), we obtain that event
\begin{equation}
\left\| {{{\bf{A}}_{\boldsymbol{{S} }}}{{({\bf{A}}_{\boldsymbol{{S} }}^{{T}}{{\bf{A}}_{\boldsymbol{{S} }}})}^{ - {{1}}}}{{\bf{W}}_{\boldsymbol{{S} }}}{{\boldsymbol{u}}_{\boldsymbol{{S} }}}}\right\|_2^2 \le \left(1+8\sqrt {\frac{{k}}{m}}\right)\frac{\xi k}{m\sigma_A^2}
\end{equation}
is satisfied with probability greater than $1-2\text{exp}(-k/2)$.

Turning to the second term in (\ref{eq2}), we have
\begin{equation}
 \left\| {({\boldsymbol{I}} - {{\bf{A}}_{\boldsymbol{{S} }}}{\bf{A}}_{\boldsymbol{{S} }}^ + )\frac{{\boldsymbol{Z}}}{{mh}}} \right\|_2^2 \le \frac{1}{mh^2}\frac{\left\|\bf Z\right\|_2^2}{m}
\end{equation}
since $({\boldsymbol{I}} - {{\bf{A}}_{\boldsymbol{{S} }}}{\bf{A}}_{\boldsymbol{{S} }}^ + )$ is an orthogonal projection matrix. On the other hand, $\left\|\boldsymbol{Z}\right\|_2^2/\sigma_Z^2$ is a $\chi^2$ variate with $m$ degrees of freedom. Thus, applying the tail bounds for $\chi^2$ variate (see Appendix J in \cite{wainwright2009sharp}), we have that for all $\epsilon \in (0,1/2)$,
\begin{equation} \label{eq6}
\mathbb{P}\left[ \left\| {({\boldsymbol{I}} - {{\bf{A}}_{\boldsymbol{{S} }}}{\bf{A}}_{\boldsymbol{{S} }}^ + )\frac{{\boldsymbol{Z}}}{{mh}}} \right\|_2^2 \geq (1+\epsilon)\frac{\sigma_Z^2}{mh^2}\right] \le \text{exp}\left(-\frac{3m\epsilon^2}{16}\right)
\end{equation}

Combining (\ref{eq2}), (\ref{eq4}) and (\ref{eq6}), we have that event
\begin{equation}
\begin{split}
&{\mathop{\rm var}} ({\Gamma _i}) \\
&\geq \Pi \buildrel \Delta \over = \sigma_A^2\left( \left(1 + \max \left\{\epsilon,\,8\sqrt {\frac{{k}}{m}} \right\}\right) \left(\frac{\xi k}{m\sigma_A^2} + \frac{{{\sigma_Z^2}}}{{m{h^2}}}\right) \right)
\end{split}
\end{equation}
is satisfied with probability less than $4\text{exp}(-c_1\text{min}\{m\epsilon^2, k\})$ for some $c_1>0$.

Consequently, applying the standard Gaussian tail bounds (see Appendix A in \cite{wainwright2009sharp}), we have
\begin{equation}\label{eq7}
\begin{split}
\mathbb{P}&\left[ {\mathop {\max }\limits_{i \in {S^c}} \left| {{\Gamma _i}} \right| \ge {w_i}} \right]\\
&= \mathbb{P}\left[ {\mathop {\max }\limits_{i \in {S^c}} \left| {{\Gamma _i}} \right| \ge {w_i}|{\rm{var}}({\Gamma _i}) < \Pi } \right]\mathbb{P}\left[ {{\rm{var}}({\Gamma _i}) < \Pi } \right]\\
&\;\;\;\;+ \mathbb{P}\left[ {\mathop {\max }\limits_{i \in {S^c}} \left| {{\Gamma _i}} \right| \ge {w_i}|{\rm{var}}({\Gamma _i}) \ge \Pi } \right]\mathbb{P}[{\rm{var}}({\Gamma _i}) \ge \Pi ]\\
&\leq\mathbb{P}\left[\mathop {\max }\limits_{i \in {{S} ^c}} \left| {{\Gamma _i}} \right| \ge w_i|{\rm{var}}({\Gamma _i}) < \Pi \right] + \mathbb{P}[{\rm{var}}({\Gamma _i}) \geq \Pi ]\\
&\leq 2(n-k)\text{exp}\left( { - \frac{w_i^2}{{2\Pi }}} \right)+4\text{exp}(-c_1\text{min}\{m\epsilon^2, k\})
\end{split}
\end{equation}
In high dimensional case, we can assume that when $m$ is sufficiently large, inequality $8\sqrt {\frac{{k}}{m}}<\epsilon$ holds for any fix $\epsilon>0$ \cite{wainwright2009sharp}. Hence, the exponential term in (\ref{eq7}) is decaying, provided
\begin{equation}
m > 2\eta k\log (n - k)(1 + \epsilon^{'})(1 + \frac{{\sigma _Z^2\sigma _A^2}}{{{h^2}k}})
\end{equation}
where $\eta=\mathop {\max}\limits_{i \in \boldsymbol{S}^c}\{\frac{\xi}{w_i^2}\}$ and $\epsilon^{'}$ satisfies $\log (n - k)(1 + \epsilon^{'})=(1+\log(n-k))(1 + \epsilon)$.

For the event 2) in (\ref{SNC3}), we establish a bound on ${\left\|\boldsymbol{{{\hat x}_{S} } - x_{S} ^*} \right\|_\infty }$. According to (\ref{eq14}) and applying triangle inequality, we have
\begin{equation}\label{eq8}
{\left\|\boldsymbol{{{\hat x}_{S} } - x_{S} ^*} \right\|_\infty }\le mh{\left\| { {{({\bf A}_{\boldsymbol{S}}^T{{\bf A}_{\boldsymbol{S}} })}}}^{ - 1}{\bf {W_{\boldsymbol{S}} }{\boldsymbol{u_{S}} }} \right\|_\infty } +\left\|\bf A^+_{\boldsymbol{S}}{\boldsymbol{Z}}\right\|_\infty
\end{equation}
For the second term in (\ref{eq8}), conditioned on $\bf A_{\boldsymbol{S}}$, random vector $\bf A^+_{\boldsymbol{S}}Z$ is zero-mean Gaussian with variance at most
\begin{equation}
\frac{{{\sigma_Z ^2}}}{m}{\left| {\left\| {{{({{\bf A}_{\boldsymbol{S}} ^T{\bf A_{\boldsymbol{S}} }}/m)}^{ - 1}}} \right\|} \right|_2}
\end{equation}
As analyzed in \cite{wainwright2009sharp},
\begin{equation}
\mathbb{P}\left[\Upsilon\buildrel \Delta \over =\frac{{{\sigma_Z ^2}}}{m}{\left| {\left\| {{{({{\bf A}_{\boldsymbol{S}} ^T{\bf A_{\boldsymbol{S}} }}/m)}^{ - 1}}} \right\|} \right|_2\geq \frac{9\sigma_Z^2}{m\sigma_A^2}} \right]\leq2\text{exp}(-m/2)
\end{equation}
By the total probability rule, it follows
\begin{equation}
\begin{split}
\mathbb{P}&\left[\left\|\bf A^+_{\boldsymbol{S}}{\boldsymbol{Z}}\right\|_\infty>t\right]\\
&\le \mathbb{P}\left[\left\|\bf A^+_{\boldsymbol{S}}{\boldsymbol{Z}}\right\|_\infty>t|\Upsilon< \frac{9\sigma_Z^2}{m\sigma_A^2}\right]+\mathbb{P}(\Upsilon\geq \frac{9\sigma_Z^2}{m\sigma_A^2})
\end{split}
\end{equation}
Using the Gaussian tail bounds (see Appendix B), it follows
\begin{equation}\label{eq10}
\mathbb{P}\left[ \left\|\bf A^+_{\boldsymbol{S}}{\boldsymbol{Z}}\right\|_\infty\geq 6\sqrt {\frac{{{\sigma_Z^2}\log (k)}}{m\sigma_A^2}} \right]\leq 4\text{exp}(-c_1m)
\end{equation}

For the first term in (\ref{eq8}), we have
\begin{equation}
\begin{split}
&mh{\left\| {{ {{({\bf A}_{\boldsymbol{S}} ^T{\bf A_{\boldsymbol{S}} })}}}^{ - 1}{\bf {W_{\boldsymbol{S}} }{\boldsymbol{u_{S}} }}} \right\|_\infty }\\
&\le h\left\{ {{{\left\| {\left[ {{{(\frac{{{\bf{A}}_{\boldsymbol{S}}^{{T}}{{\bf{A}}_{\boldsymbol{S}}}}}{m})}^{ - {{1}}}} - {{\boldsymbol{I}}^{ - 1}}} \right]{{\bf{W}}_{\boldsymbol{S}}}{{\boldsymbol{u}}_{\boldsymbol{S}}}} \right\|}_\infty } + {{\left\| {{{\boldsymbol{I}}^{ - 1}}{{\bf{W}}_{\boldsymbol{S}}}{{\boldsymbol{u}}_{\boldsymbol{S}}}} \right\|}_\infty }} \right\}
\end{split}
\end{equation}
According to Lemma 5 in \cite{wainwright2009sharp}, we have
\begin{equation}
\begin{split}
\mathbb{P}&\left\{ {{{\left\| {\left[ {{{(\frac{{{\bf{A}}_{\boldsymbol{S}}^{{T}}{{\bf{A}}_{\boldsymbol{S}}}}}{m})}^{ - {{1}}}} - {{\boldsymbol{I}}^{ - 1}}} \right]{{\bf{W}}_{\boldsymbol{S}}}{{\boldsymbol{u}}_{\boldsymbol{S}}}} \right\|}_\infty } \ge {c_1}{{\left\| {{{\bf{W}}_{\boldsymbol{S}}}{{\boldsymbol{u}}_{\boldsymbol{S}}}} \right\|}_\infty }} \right\}\\
&\le 4\exp ( - {c_2}\min \{ k,\log (n - k)\} )
\end{split}
\end{equation}
holds for some $c_2>0$.
Therefore, it follows
\begin{equation}\label{eq11}
\begin{split}
\mathbb{P}&\left\{ {mh{{\left\| {{{({\bf{A}}_{\boldsymbol{S}}^{{T}}{{\bf{A}}_{\boldsymbol{S}}})}^{ - {{1}}}}{{\bf{W}}_{\boldsymbol{S}}}{{\boldsymbol{u}}_{\boldsymbol{S}}}} \right\|}_\infty } > {c_3}h{{\left\| {{{\bf{W}}_{\boldsymbol{S}}}{{\boldsymbol{u}}_{\boldsymbol{S}}}} \right\|}_\infty }} \right\}\\
&\le c_3^{'}\exp ( - {c_2}\min \{ k,\log (n - k)\} )
\end{split}
\end{equation}
holds for some $c_3,\;c_3^{'}>0$.

Combining (\ref{eq8}), (\ref{eq10}) and (\ref{eq11}), we have that event
\begin{equation}
{\left\|\boldsymbol{{{\hat x}_{S} } - x_{S} ^*} \right\|_\infty }\le {c_3}h{\left\| {{{\bf{W}}_{\boldsymbol{S}}}{{\boldsymbol{ u}}_{\boldsymbol{S}}}} \right\|_\infty } + 6\sqrt {\frac{{\sigma _Z^2\log (k)}}{{m\sigma _A^2}}}\buildrel \Delta \over =g(h)
\end{equation}
is satisfied with probability greater than $1 - {c_3^{'}}\exp ( - {c_2}\min \{ k,\log (n - k)\})$.

Therefore, if $\forall i \in \boldsymbol{S}$ $|x_i^*|>g(h)$ holds, we have that for all $i \in \boldsymbol{S}$, $\mbox{sign}(\hat x_i)=\mbox{sign}(x^*_i)$ hold with high probability. Combining the probabilities that two events in (\ref{SNC3}) are satisfied, the conclusion in Theorem \ref{recoverytheorem} holds.

\bibliographystyle{IEEEtran}

\end{document}